\documentclass[12pt,british]{mhd}
\usepackage[T1]{fontenc}
\usepackage[latin9]{inputenc}
\usepackage{geometry}
\usepackage{units}
\usepackage{amsmath}
\usepackage{amssymb}
\usepackage{graphicx}
\usepackage{setspace}
\usepackage{esint}
\makeatletter
\def\Rm{\mathit{Rm}}

\renewcommand{\vec}[1]{\mbox{\boldmath $#1$}}

\usepackage{balance}

\usepackage{babel}

\mhdhead{59}{2}{1}

\makeatother

\usepackage{babel}
\begin{document}
\title{Optimizing helical disc dynamo}
\author{J. Priede\inst{1,2},R. A. Avalos-Zúñiga\inst{3}}
\institute{Fluid and Complex Systems Research Centre, Coventry University,\\Coventry, CV1 5FB, United Kingdom \and Department of Physics, University of Latvia, Riga, LV-1004, Latvia
\and  Research Centre in Applied Science and Advanced technology (CICATA - Querétaro),
Instituto Politécnico Nacional, Cerro Blanco 141, Colinas del Cimatario, Querétaro, Mexico}
\maketitle
\begin{abstract}
We present an optimized design of our recently realized helical disc dynamo.
Like the original set-up, the optimized dynamo consists of a flat
multi-arm spiral coil and a co-axially placed disc which is connected
to the former by sliding liquid metal contacts. In contrast to the
original set-up, the disc and the coil in the optimized design have
different sizes. This allows the disc to capture more of the high-density
magnetic flux generated in the inner part of the coil and to avoid
the reverse flux in the outer part of the coil. By optimizing the
coil and dics radii, the critical magnetic Reynolds number can be
reduced from 
\global\long\def\Rm{\mathit{Rm}}%
$\approx34.6$ when the disc and coil have equal inner and outer radii
with the ratio $r_{i}/r_{o}\approx0.36$ to $\Rm\approx11.6.$ This
lowest possible disc dynamo threshold is attained when the disc and
coil have relatively narrow widths. Using a slightly suboptimal but
more practical set-up with the inner and outer radii of the disc and
coil equal to to $(0.3,0.9)$ and $(0.74,1),$ respectively, self-excitation
is expected at $\Rm\approx14.6.$ 

\end{abstract}

\section{Introduction.}

The disc dynamo is a very simple physical model that is often used
to illustrate the self-excitation of the magnetic field by moving
conductors \cite{Moffatt1978}. This is known as the dynamo action, which is currently 
the most likely explanation for the origin of the
magnetic fields of the Earth, the Sun, and other cosmic bodies \cite{Beck1996}.
The basic disc dynamo model, which is due to Bullard \cite{Bullard1955},
consists of a solid metal disc that rotates about its central
axis, and a wire that is twisted around and connected through sliding
contacts to the rim and the axis of the disc. If the disc spins sufficiently
fast and in the right direction, such a set-up can amplify electric
current circulating in the system and, thus, the associated magnetic
field. This happens when the rotation rate of the disc exceeds a certain
critical threshold above which the potential difference induced across
the disc exceeds the voltage drop caused by the ohmic resistance of
the system. Then the current starts to grow exponentially in time, 
resulting in the self-excitation of the magnetic field. The growth
stops when the braking electromagnetic torque becomes so strong that it slows the disc.

Bullard dynamo has several extensions and modifications. The most
well-known is perhaps the so-called Rikitake dynamo \cite{Rikitake1958},
which consists of two coupled disc dynamos. This dynamo can generate
an oscillating magnetic field with complex dynamics \cite{Plunian1998}.
The latter study uses a modified disc dynamo model proposed by \cite{Moffatt1979}.
In this model, the disc is radially segmented, and azimuthal current
is enforced at the rim. The aim of these modifications is to eliminate
the unphysical growth of the magnetic field in the limit of perfectly
conducting disc. We used this model to show that the dynamo can be excited
by the parametric resonance mechanism at a substantially reduced rotation
rate when the latter contains harmonic oscillations in certain frequency
bands \cite{Priede2010}.

Despite its simplicity, the implementation of the disc dynamo is faced
with severe technical difficulties. The most challenging problem is
the sliding electrical contacts which are required to convey the current
between the rim and the axis of the rotating disc. The electrical
resistance of such contacts, which are usually made of solid graphite
brushes, can significantly exceed that of the rest of the set-up.
This results in an unrealistically high rotation rate required for
the dynamo to operate. Therefore, in contrast to fluid dynamos,
which have been performed in several laboratory experiments using liquid
metal \cite{Gailitis2001,Stieglitz2001,Monchaux2007} (\cite{Stefani2019}
in preparation), the disc dynamo was generally considered technically
unfeasible \cite{Raedler2002,Dormy2007,Lorrain2007}.

To overcome this problem, we proposed a feasible disc dynamo design
with sliding liquid-metal electrical contacts \cite{Priede2013}.
This design was realized in a set-up consisting of a copper disc with
a radius of $\unit[30]{cm}$ and thickness of $\unit[3]{cm}$ which
was placed coaxially beneath a flat multi-arm spiral coil of the
same size and electrically connected to it electrically at the centre and along the circumference by sliding liquid-metal contacts \cite{Avalos-Zuniga2017}.
The slits make the conductivity of the coil anisotropic, which allows this essentially axially symmetric dynamo to generate an axially symmetric
magnetic field \cite{Plunian2020,Alboussiere2022}. In this set-up, we observed a dynamo action when the disc rotation rate reached $\Omega\approx\unit[7]{Hz}$
\cite{Avalos-Zuniga2023}. 

It is important to note that this observed critical rotation rate
is somewhat lower than the theoretically predicted threshold $\Omega_{c}\approx\unit[8.2]{Hz}$
which corresponds to the minimal critical magnetic Reynolds number
$\Rm\approx35$ for a negligible contact resistance \cite{Priede2013}.
The most likely reason for this difference is the very approximate
nature of the current sheet model, which we used to evaluate the dynamo
threshold \cite{Priede2013}. In particular, our original disc dynamo
model is based on the assumption that the magnetic flux through the
disc is the same as that through the coil. In reality, however, the
effective inner radius of the disc is somewhat smaller than that of
the coil. Due to the relatively high strength of the magnetic field
in the central part of the helical coil, the total magnetic flux through
the disc may be larger than theoretically predicted. This hints at
the possibility, which is investigated theoretically in the present
note, that the critical rotation rate can be reduced further by optimizing
the inner and outer radii of the disc and coil.

\section{Mathematical model.}

\begin{figure}
\noindent \begin{centering}
\includegraphics[bb=0bp 0bp 739bp 331.009bp,width=0.5\columnwidth]{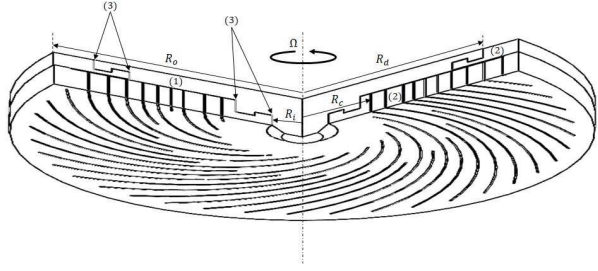}
\par\end{centering}
\caption{\label{fig:sketch}Schematic bottom view of the set-up which consists
of a coil (1) and a disc (2) with inner radius $r_{i}$ and the outer
radius $r_{d}.$ The coil is sectioned by spiral slits, which extend
from the inner radius $r_{c}$ to the outer radius $r_{o},$ and has
two solid parts: the lower one between radii $r_{i}$ and $r_{c}$
and the upper one between $r_{d}$ and $r_{o},$ which are electrically
connected to the disc by sliding liquid-metal contacts (3).}
\end{figure}

The principal set-up of the disc dynamo considered in this study is
shown in Fig. \ref{fig:sketch}. It consists of a thin rotating disc with inner radius $r_{i}$ and outer radius $r_{d},$ and
a coaxial coil of similar shape with the same inner radius $r_{i}$
but a larger outer radius $r_{o}>r_{d}.$ In contrast to our original
set-up \cite{Priede2013}, now the coil is placed below the disc and
sectioned by spiral slits starting only from the radius $r_{c}>r_{i}.$
The sectioned part of the coil, which generates an axial magnetic field,
extends to the outer radius $r_{o}.$ Between the radii $r_{d}$ and
$r_{o},$ the coil has a solid upper part that connects its rim to
that of the disc by a sliding liquid-metal contact. The circuit made
of the disc and the coil is closed by the second sliding liquid-metal
contact which is located at the inner radius $r_{i}.$ 

As before, to simplify the analysis, the thickness of the disc and
the coil is assumed to have the same thickness $d$, which is much smaller
than the radial dimensions of the system. The same assumption applies
also to the vertical separation between the disc and the coil. Thus, the disc and
the coil are treated as vertically collocated thin sheets with effective linear conductance $\bar{\sigma}=\sigma d.$ 

Now consider an equilibrium dynamo state in which the disc rotating
with a fixed angular velocity $\Omega$ in the magnetic field $\vec{B}$
generated by a constant electric current $I_{0}$ sustains this current.
In the solid disc this stationary current flows is purely radial with
the linear density $J_{r}=\frac{I_{0}}{2\pi r},$ which decreases
due to charge conservation inversely with the cylindrical radius
$r.$ This current flowing in the opposite radial direction through
the coil is deflected sideways by the spiral slits, which thus produce
an azimuthal current component proportional to the radial one:

\begin{equation}
J_{\phi}=-J_{r}\beta=\frac{I_{0}\beta}{2\pi r},\label{eq:J-phi}
\end{equation}
where $\arctan\beta$ the pitch angle of the current lines relative
to the radial direction. The shape of slits is governed by $\frac{J_{\phi}}{J_{r}}=\frac{rd\phi}{dr}=-\beta.$
The general solution of this ordinary differential equation 
\begin{equation}
\phi(r)=\phi_{0}-\beta\ln r\label{eq:spiral}
\end{equation}
describes logarithmic spirals. The electric potential distribution
in the sectioned part of the coil can be found from the Ohm's law
\begin{equation}
\vec{J}=\frac{I_{0}}{2\pi r}\left(-\vec{e}_{r}+\beta\vec{e}_{\phi}\right)=-\sigma d\vec{\nabla}\varphi,\label{eq:Jc}
\end{equation}
as 
\[
\varphi(r,\phi)=\frac{I_{0}}{2\pi\sigma d}\left(\ln r-\beta\phi\right).
\]
Thus, the potential difference along the current line over the sectioned
part of the coil is 
\begin{equation}
\Delta\varphi_{c}^{o}=\left[\varphi(r,\phi(r))\right]_{r_{c}}^{r_{o}}=\frac{I_{0}}{2\pi\sigma d}(1+\beta^{2})\ln\frac{r_{o}}{r_{c}}.\label{eq:Phi-c}
\end{equation}
Over the solid part of the coil $(r_{i}\le r\le r_{c}),$ where the
current is purely radial, which corresponds to $\beta=0,$ we respectively
have
\[
\Delta\varphi_{i}^{c}=\frac{I_{0}}{2\pi\sigma d}\ln\frac{r_{c}}{r_{i}}.
\]
The potential difference across the disc, which rotates as a solid
body with the azimuthal velocity $v_{\phi}=r\Omega,$ is defined by
the radial component of Ohm's law for a moving medium 
\[
J_{r}=\frac{I_{0}}{2\pi r}=\sigma d(-\partial_{r}\varphi+v_{\phi}B_{z}),
\]
where $B_{z}$ is the axial component of the magnetic field. Integrating
the expression above over the upper part of the set-up from $r_{i}$
to $r_{o}$disc, which includes the disc and the solid upper part
of the coil, where $\Omega=0,$ we have 
\begin{equation}
\frac{I_{0}}{2\pi}\ln\frac{r_{o}}{r_{i}}=\sigma d\left(-\Delta\varphi_{i}^{o}+\Omega\Phi_{i}^{d}\right),\label{eq:Phi-d}
\end{equation}
where $\Delta\varphi_{i}^{o}=\left[\varphi_{d}(r)\right]_{r_{i}}^{r_{o}}$
is the potential difference across the upper part of the set-up and
$\Phi_{i}^{d}=\int_{r_{i}}^{r_{d}}B_{z}r\,dr$ is the magnetic flux
through the disc. Using the relation $B_{z}=r^{-1}\partial_{r}(rA_{\phi}),$
the latter can be expressed in terms of the azimuthal component of
the magnetic vector potential $A_{\phi}$ as 
\begin{equation}
\Phi_{i}^{d}=\left[rA_{\phi}\right]_{r=r_{i}}^{r_{d}}.\label{eq:flux-d}
\end{equation}
In the stationary state, which is assumed here, the potential difference
induced by the rotating ring in Eq. (\ref{eq:Phi-d}) is supposed
to balance that over the coil defined by Eq. (\ref{eq:Phi-c}) as
well as the potential drop over the liquid-metal contacts with the
effective resistance $\mathcal{R}:$ 
\begin{equation}
\Delta\varphi_{i}^{o}=\Delta\varphi_{i}^{c}+\Delta\varphi_{c}^{o}+\mathcal{R}I_{0}.\label{eq:blnc}
\end{equation}
This equation implicitly defines the critical rotation rate at which
a steady current can sustain itself.

To complete the solution, we need to evaluate the magnetic flux (\ref{eq:flux-d})
through the rotating disc. The azimuthal component of the vector potential that appears in Eq. (\ref{eq:flux-d}) is generated by the respective
component of the electric current which is present only in the coil.
Thus, we have 
\[
A_{\phi}(r,z)=\frac{\mu_{0}}{4\pi}\intop_{0}^{2\pi}\intop_{r_{c}}^{r_{o}}\frac{J_{\phi}(r')\cos\phi\,r'\,dr'\,d\phi}{\sqrt{r'^{2}-2r'r\cos\phi+r^{2}+z^{2}}},
\]
where $z$ is the axial distance from the coil ring carrying the azimuthal
current $J_{\phi}$ defined by Eq. (\ref{eq:J-phi}). Note that the
poloidal currents with radial and axial components circulating through
the rings and liquid-metal contacts produce a purely toroidal magnetic
field, which is parallel to the velocity of the rotating ring, and thus do not interact with the latter. In the plane of the ring $(z=0),$
the double integral above can be evaluated analytically as 
\[
A_{\phi}(r,0)=\frac{\mu_{0}\beta I_{0}}{8\pi^{2}}\left[F(r_{c}/r)-F(r_{o}/r)\right],
\]
where $F(x)=(1-x)K(m_{+})+(1+x)E(m_{+})+\mathrm{sgn}(1-x)\left[(1+x)K(m_{-})+(1-x)E(m_{-})\right]$
is defined in terms of the complete elliptic integrals of the first
and second kind, $K(m_{\pm})$ and $E(m_{\pm})$, of the \emph{parameter}
$m_{\pm}=\frac{\pm4x}{(1\pm x)^{2}}$ \cite{Abramowitz1972}. Using
the above definition, the magnetic flux (\ref{eq:flux-d}) can be
written as 
\begin{equation}
\Phi_{i}^{d}=\frac{\mu_{0}\beta I_{0}r_{o}}{8\pi^{2}}\bar{\Phi}(\lambda_{i},\lambda_{c},\lambda_{d})\label{eq:flux-c}
\end{equation}
where 
\begin{equation}
\bar{\Phi}(\lambda_{i},\lambda_{c},\lambda_{d})=\lambda_{d}\left[F(\lambda_{c}/\lambda_{d})-F(\lambda_{d}^{-1})\right]-\lambda_{i}\left[F(\lambda_{c}/\lambda_{i})-F(\lambda_{i}^{-1})\right]\label{eq:flux-l}
\end{equation}
is a dimensionless magnetic flux and $\lambda_{i}=r_{i}/r_{o},$$\lambda_{c}=r_{c}/r_{o},$$\lambda_{d}=r_{d}/r_{o}$
are the dimensionless counterparts of the corresponding radii which
are defined using the outer coil radius $r_{o}$ as the length scale
for this problem. Substituting the relevant parameters into Eq. (\ref{eq:blnc}),
we obtain the critical magnetic Reynolds number
\begin{equation}
\Rm=\mu_{0}\sigma dr_{o}\Omega=\frac{4\pi(\bar{\mathcal{R}}-2\ln\lambda_{i}-\beta^{2}\ln\lambda_{c})}{\beta\bar{\Phi}(\lambda_{i},\lambda_{c},\lambda_{d})},\label{eq:Rm}
\end{equation}
which defines the dynamo threshold depending on the dimensionless
contact resistance $\bar{\mathcal{R}}=2\pi\sigma d\mathcal{R},$ the
spiral pitch $\beta,$ and three radii ratios $\lambda_{i},\lambda_{c},\lambda_{d}.$

\section{Results}

In the following, we assume $\bar{\mathcal{R}}=0,$ which corresponds
to a relatively small contact resistance, as in our previous set-up,
and search for an optimal set the remaining four parameters which
yield the lowest possible $\Rm.$ First, an optimal $\beta$ is found
in terms of the the radii ratios by solving the stationarity condition
$\frac{\partial\Rm}{\partial\beta}=0,$ which reduces to 
\begin{equation}
2\beta^{-2}\ln\lambda_{i}-\ln\lambda_{c}=0,
\end{equation}
and straightforwardly yields 
\begin{equation}
\beta=\sqrt{2\ln\lambda_{i}/\ln\lambda_{c}}.\label{eq:beta}
\end{equation}
Substituting this back into Eq. (\ref{eq:Rm}), we obtain 
\begin{equation}
\Rm=\frac{16\pi}{\sqrt{2}}\frac{\sqrt{\ln\lambda_{i}\ln\lambda_{c}}}{\bar{\Phi}(\lambda_{i},\lambda_{c},\lambda_{d})}
\end{equation}
depending on $\lambda_{i},\lambda_{c},\lambda_{d}.$ 

\begin{figure}
\begin{centering}
\includegraphics[width=0.5\columnwidth]{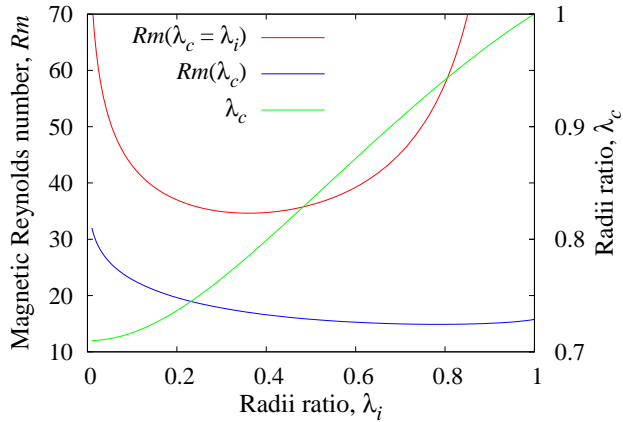} 
\par\end{centering}
\caption{\label{fig:Rm-R0}Marginal $\protect\Rm$ versus the inner disc radius
$\lambda_{i}$ for the same-size inner coil radius $\lambda_{c}=\lambda_{i}$
and for optimal $\lambda_{c}$ which is shown on the axis at the right;
here $\lambda_{d}=1$ and $\bar{\mathcal{R}}=0.$ }
\end{figure}

\begin{figure}
\begin{centering}
\includegraphics[width=0.5\columnwidth]{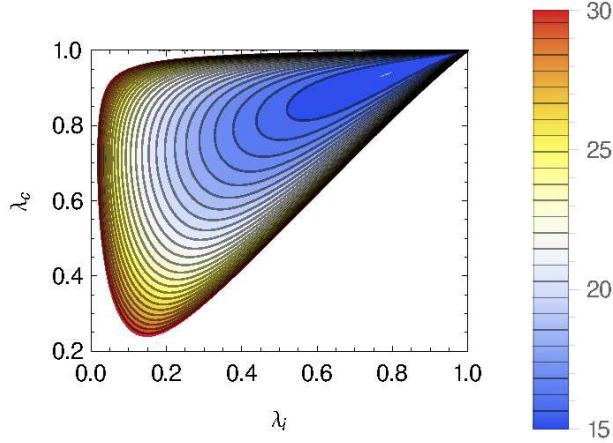}
\par\end{centering}
\caption{\label{fig:Rm-ci} Isocontours of the magnetic Reynolds number $\protect\Rm$
in the plane of radii ratios $(\lambda_{i},\lambda_{c})$ for $\lambda_{i}\le\lambda_{c}$
and $\lambda_{d}=1.$ }
\end{figure}

With $\lambda_{d}=1$ and $\lambda_{i}=\lambda_{c},$ which correspond
to $r_{c}=r_{i}$ and $r_{d}=r_{o}$, as in our original model, the
lowest $\Rm\approx34.6$ is obtained in the radii ratio $\lambda=r_{i}/r_{o}\approx0.36$
(see Fig. \ref{fig:Rm-R0}) \cite{Priede2013}. First, this basic
set-up can be improved by optimazing the inner coil radius $\lambda_{c}$
so that to minimize $\Rm$ for the given inner disc radius $\lambda_{i}$.
The optimal $\lambda_{c}$ along with the corresponding $\Rm$ found
numerically using Mathematica \cite{Wolfram2003} are plotted in Fig.
\ref{fig:Rm-R0} against $\lambda_{i}.$ As seen in Fig. \ref{fig:Rm-ci},
for this partially constrained set-up with $\lambda_{d}=1,$ the lowest
$\Rm\approx14.9$ is attained at $\lambda_{i}\approx0.78$ and $\lambda_{c}\approx0.94.$
Optimizing also $\lambda_{d},$ we find the lowest possible $\Rm\approx11.6$
at $\lambda_{d}\approx0.98,$ $\lambda_{i}\approx0.88$ and $\lambda_{c}\approx0.96.$ 

It is important to note that the coil in this optimal set-up represents
a very narrow ring with the relative width $1-\lambda_{c}\approx0.04.$
First, such a coil may be difficult to make. Second, it may not
be adequately described by the thin-sheet approximation used in our
mathematical model. The same applies largely also to the previous (suboptimal) set-up with $\lambda_{d}=1.$ On the other hand, Fig.
\ref{fig:Rm-R0} indicates that the increase in $\Rm$ caused by the
reduction of $\lambda_{i}$ below its optimal value remains relatively
small down to $\lambda_{i}\approx0.3.$ Choosing this value for the
inner disc radius, and minimizing $\Rm$ with respect to $\lambda_{c}$
and $\lambda_{d},$ we obtain $\Rm\approx14.6$ at $\lambda_{c}\approx0.74$
and $\lambda_{d}\approx0.9$. The spiral pitch angle, which corresponds to the optimal $\beta$ and is defined by Eq. (\ref{eq:beta}), is $\arctan\beta\approx70.5^{\circ}.$ 

\section{Conclusion.}

We found a feasible disc dynamo set-up with a critical magnetic Reynolds
number $\Rm\approx15$ that is more than twice lower than that for
our original design. For a set-up with an outer radius $R_{o}=\unit[15]{cm}$
and a thickness of $d=\unit[1.5]{cm,}$ that is twice smaller than
in our original set-up, dynamo is expected to work at the critical
rotation frequency 
\[
f=\frac{\Omega}{2\pi}=\frac{\Rm}{2\pi\mu_{0}\sigma_{\mathrm{Cu}}dR_{o}}\approx\unit[15]{Hz},
\]
where $\mu_{0}=\unit[4\pi\times10^{-7}]{H/m}$ is the vacuum permeability and $\sigma_{\mathrm{Cu}}=\unit[5.96\times10^{7}]{S/m}$
is the conductivity of copper. For sliding contacts with a gap of
width $\unit[\delta=0.25]{mm}$ filled with the eutectic alloy of
GaInSn, as in our original set-up \cite{Avalos-Zuniga2023}, the relative
contact resistance can be estimated as \cite{Priede2013}

\[
\bar{\mathcal{R}}\approx\frac{\sigma_{\mathrm{Cu}}}{\sigma_{\mathrm{GaInSn}}}\frac{\delta}{R_{o}}\left(\frac{1}{\lambda_{i}}+\frac{1}{\lambda_{d}}\right)\approx10^{-2},
\]
where $\sigma_{\mathrm{GaInSn}}=\unit[3.3\times10^{6}]{S/m}$ \cite{Mueller2001}.
This confirms the assumed negligibility of $\bar{\mathcal{R}}.$

\bibliographystyle{plain}
\bibliography{dynamo}

\begin{thebibliography}{10}

\bibitem{Abramowitz1972}
M.~Abramowitz and I.~A. Stegun.
\newblock {\em Handbook of Mathematical Functions}.
\newblock Dover, New York, 1972.

\bibitem{Alboussiere2022}
T.~Alboussi{\`e}re, F.~Plunian, and M.~Moulin.
\newblock Fury: an experimental dynamo with anisotropic electrical
  conductivity.
\newblock {\em Proc. R. Soc. Lond. A}, 478:20220374, 2022.

\bibitem{Avalos-Zuniga2023}
A.~R. Avalos-Z{\'u}{\~n}iga and J.~Priede.
\newblock Realization of {B}ullard's disc dynamo.
\newblock {\em Proc. R. Soc. Lond. A}, 479(2271):20220740, 2023.

\bibitem{Avalos-Zuniga2017}
R.A. Avalos-Z{\'u}{\~n}iga, J~Priede, and CE~Bello-Morales.
\newblock A homopolar disc dynamo experiment with liquid metal contacts.
\newblock {\em Magnetohydrodynamics}, 53:341--348, 2017.

\bibitem{Beck1996}
R.~Beck, A.~Brandenburg, D.~Moss, A.~Shukurov, and D.~Sokoloff.
\newblock Galactic magnetism: recent developments and perspectives.
\newblock {\em Annu. Rev. Astron. Astrophys.}, 34(1):155--206, 1996.

\bibitem{Bullard1955}
E.~Bullard.
\newblock The stability of a homopolar dynamo.
\newblock {\em Proc. Camb. Phil. Soc.}, 51(4):744--760, 1955.

\bibitem{Dormy2007}
E.~Dormy and A.~M. Soward.
\newblock {\em Mathematical aspects of natural dynamos}.
\newblock CRC, 2007.

\bibitem{Gailitis2001}
A.~Gailitis, O.~Lielausis, E.~Platacis, S.~Dement'ev, A.~Cifersons, G.~Gerbeth,
  T.~Gundrum, F.~Stefani, M.~Christen, and G.~Will.
\newblock Magnetic field saturation in the riga dynamo experiment.
\newblock {\em Phys. Rev. Lett.}, 86(14):3024, 2001.

\bibitem{Lorrain2007}
P.~Lorrain, F.~Lorrain, and S.~Houle.
\newblock {\em Magneto-fluid dynamics: fundamentals and case studies of natural
  phenomena}.
\newblock Springer, 2006.

\bibitem{Moffatt1978}
H.~K. Moffatt.
\newblock {\em Magnetic field generation in electrically conducting fluids}.
\newblock Cambridge monographs on mechanics and applied mathematics. Cambrige,
  1978.

\bibitem{Moffatt1979}
H.~K. Moffatt.
\newblock A self-consistent treatment of simple dynamo systems.
\newblock {\em Geophys. Astrophys. Fluid Dyn.}, 14(1):147--166, 1979.

\bibitem{Monchaux2007}
R.~Monchaux, M.~Berhanu, M.~Bourgoin, M.~Moulin, Ph. Odier, J.-F. Pinton,
  R.~Volk, S.~Fauve, N.~Mordant, F.~P\'etr\'elis, A.~Chiffaudel, F.~Daviaud,
  B.~Dubrulle, C.~Gasquet, L.~Mari\'e, and F.~Ravelet.
\newblock Generation of a magnetic field by dynamo action in a turbulent flow
  of liquid sodium.
\newblock {\em Phys. Rev. Lett.}, 98:044502, Jan 2007.

\bibitem{Mueller2001}
U.~M{\"u}ller and L.~B{\"u}hler.
\newblock {\em Magnetofluiddynamics in Channels and Containers}.
\newblock Springer, 2001.

\bibitem{Plunian2020}
F.~Plunian and T.~Alboussi{\`e}re.
\newblock Axisymmetric dynamo action is possible with anisotropic conductivity.
\newblock {\em Phys. Rev. Res.}, 2(1):013321, 2020.

\bibitem{Plunian1998}
F.~Plunian, Ph. Marty, and A.~Alemany.
\newblock Chaotic behaviour of the rikitake dynamo with symmetric mechanical
  friction and azimuthal currents.
\newblock {\em Proc. R. Soc. Lond. A}, 454(1975):1835--1842, 1998.

\bibitem{Priede2013}
J.~Priede and R.~Avalos-Z{\'u}{\~n}iga.
\newblock Feasible homopolar dynamo with sliding liquid-metal contacts.
\newblock {\em Phys. Lett. A}, 377(34-36):2093--2096, 2013.

\bibitem{Priede2010}
J.~Priede, R.~Avalos-Z{\'u}{\~n}iga, and F.~Plunian.
\newblock Homopolar oscillating-disc dynamo driven by parametric resonance.
\newblock {\em Phys. Lett. A}, 374(4):584--587, 2010.

\bibitem{Raedler2002}
K.H. R{\"a}dler and M.~Rheinhardt.
\newblock Can a disc dynamo work in the laboratory?
\newblock {\em Magnetohydrodynamics}, 38(1-2):211--217, 2002.

\bibitem{Rikitake1958}
T.~Rikitake.
\newblock Oscillations of a system of disk dynamos.
\newblock {\em Math. Proc. Cambridge Philos. Soc.}, 54(1):89--105, 1958.

\bibitem{Stefani2019}
F.~Stefani, A.~Gailitis, G.~Gerbeth, A.~Giesecke, Th. Gundrum, G.~R{\"u}diger,
  M.~Seilmayer, and T.~Vogt.
\newblock The dresdyn project: liquid metal experiments on dynamo action and
  magnetorotational instability.
\newblock {\em Geophys. Astrophys. Fluid Dyn.}, 113(1-2):51--70, 2019.

\bibitem{Stieglitz2001}
R.~Stieglitz and U.~M{\"u}ller.
\newblock Experimental demonstration of a homogeneous two-scale dynamo.
\newblock {\em Phys. Fluids}, 13(3):561--564, 2001.

\bibitem{Wolfram2003}
S.~Wolfram.
\newblock {\em The Mathematica Book}.
\newblock Wolfram Media, 2003.

\end{thebibliography}

\lastpageno
\end{document}